\documentclass[conference]{IEEEtran}
\IEEEoverridecommandlockouts
\usepackage{cite}
\usepackage{xurl}
\usepackage{hyperref}
\usepackage{amsmath,amssymb,amsfonts}
\usepackage{algorithmic}
\usepackage{graphicx}
\usepackage{textcomp}
\usepackage{xcolor}
\usepackage{stfloats}
\usepackage{balance}
\def\BibTeX{{\rm B\kern-.05em{\sc i\kern-.025em b}\kern-.08em
    T\kern-.1667em\lower.7ex\hbox{E}\kern-.125emX}}
\begin{document}

\title{Virtual Microphone Array for Direction-of-Arrival Estimation Based on Fiber-Optic Position Sensing
\vspace{1em}
}

\author{
\IEEEauthorblockN{
Sebastian Schwarze\IEEEauthorrefmark{1},
Jiekai Wang\IEEEauthorrefmark{2},
Thomas Kissinger\IEEEauthorrefmark{2},
Stephan Werner\IEEEauthorrefmark{1}
}
\IEEEauthorblockA{
\textit{Technische Universität Ilmenau, Germany}\\
\IEEEauthorrefmark{1}\textit{Electronic Media Technology Group}\\
\IEEEauthorrefmark{2}\textit{Nanofabrication and Nanomeasurement Technology Group}\\
Corresponding author: sebastian.schwarze@tu-ilmenau.de
}

}

\maketitle

\begin{abstract}
This paper presents a virtual microphone array approach for estimating the direction of arrival of a sound source using a moving microphone and a stationary reference microphone. The method exploits the microphone motion to generate a time-varying array geometry, using fiber-optic position sensing to reconstruct the trajectory and determine the resulting virtual microphone configurations. For each configuration, the estimated time difference of arrival between the two microphone signals and the corresponding microphone geometry define a hyperboloid representing all possible sound source positions consistent with the time difference. By combining multiple hyperboloid constraints, the direction of arrival can be estimated in azimuth and elevation relative to the microphone coordinate system. In contrast to conventional arrays with fixed geometries and approaches based on predefined microphone trajectories, the proposed method derives the virtual array geometry directly from the measured microphone motion, enabling a spatially distributed virtual array to be formed using only two physical microphones and offering potential for size- and weight-constrained applications. The approach was experimentally evaluated under controlled low-reflection conditions for different source positions, source distances, and signal characteristics. The influence of microphone motion was further examined, and the method was tested with a moving sound source. The results demonstrate the feasibility of direction-of-arrival estimation using a virtual microphone array without requiring a predefined microphone trajectory.
\end{abstract}

\vspace{1em}

\begin{IEEEkeywords}
Virtual Microphone Array, Direction-of-Arrival Estimation, Fiber-Optic Position Sensing
\end{IEEEkeywords}

\vspace{1em}

\section{Introduction}
Determining the direction of sound sources is a fundamental task in numerous audio engineering applications and systems. The information can be used, for example, to enhance desired signal components, suppress unwanted interference, or localize noise sources \cite{microphone_array}. Such methods are used in applications such as hearing aids, conference systems, acoustic cameras, and drone-based sound source localization \cite{UAV_SSL}.
To extract spatial information from the sound field, microphone arrays with predefined geometries and a fixed number of microphones are typically employed. The performance of direction estimation is influenced by the array aperture as well as by the number and spatial arrangement of the microphones \cite{mircophone_array_signal_processing}.
To avoid spatial aliasing when processing broadband signals, dense spatial sampling of the sound field is required. This typically necessitates a large number of microphones and therefore increases the technical complexity of array design, synchronization, calibration, and data processing \cite{aliasing_in_microphone_arrays}.

One approach to reducing the required number of sensors is to exploit the motion of one or more microphones. By analyzing phase or frequency changes in the recorded signals, information about the direction of a sound source relative to the microphones can be obtained \cite{rotating_array,7bell}. Existing systems typically achieve this by moving the microphones along predefined trajectories. While this simplifies the determination of microphone positions, it constrains the resulting virtual array geometry, which can complicate the implementation of such systems in applications with strict size and weight constraints.
An alternative is to derive the virtual array geometry from non-predefined microphone motion. By temporally segmenting the recorded signals and associating each segment with the corresponding microphone position, a virtual microphone array comprising multiple spatially distributed microphone positions can be constructed \cite{virtual_arrays}. The resulting sequential measurements provide relative propagation-time information between different spatial positions that can be used for direction-of-arrival estimation. However, this approach places stringent requirements on the position sensing system, since the microphone trajectory must be determined with high spatial accuracy and sufficient temporal resolution.

\begin{figure}[!h]
    \begin{center}
        \includegraphics[width=\columnwidth]{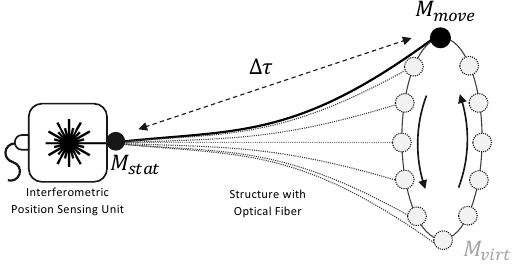}
    \end{center}
    \caption{\textbf{Concept of the proposed virtual microphone array.} The motion of the supporting structure carrying the microphone $M_\text{move}$ generates a sequence of virtual microphone positions $M_\text{virt}$. The Interferometric Position Sensing Unit uses fiber-optic shape sensing to reconstruct the microphone trajectory and determine the corresponding virtual microphone positions. For each $M_\text{virt}$, the time difference of arrival (TDOA) $\Delta\tau$ relative to the stationary reference microphone $M_\text{stat}$ is estimated. Combining the resulting microphone configurations and TDOAs enables DOA estimation of the sound source.
    }
    \label{fig:Concept_Einleitung}
\end{figure}

A promising approach to addressing these requirements is the use of fiber-optic interferometric sensing. Such systems enable highly accurate measurements of the displacement of mechanical structures and are used in a wide range of technical applications, including vibration analysis, manufacturing metrology, and geophysical structural analysis \cite{Fiber_optic_Displacement}. The measured displacement can further be used to derive kinematic quantities and to reconstruct the motion of mechanical structures.
Combining virtual-microphone-array-based direction-of-arrival estimation with fiber-optic position sensing for determining the virtual microphone positions forms the basis of the approach presented in this paper (Fig.~\ref{fig:Concept_Einleitung}).

\section{Background}
\subsection{Direction-of-Arrival Estimation}
\label{sec:DOA}

Direction-of-arrival (DOA) estimation refers to determining the direction from which an acoustic or electromagnetic wavefront reaches a sensor system, such as an antenna or microphone array \cite{optimum_array_processing}. Various methods can be employed for this purpose, and their suitability depends on the geometry of the array and the prevailing environmental conditions. Common approaches include correlation- and time-difference-based methods, delay-and-sum beamforming, and subspace-based methods \cite{localization_in_robotics}.

The most fundamental approach is based on the analysis of the time difference of arrival (TDOA) of a wavefront between at least two microphones. If the microphone spacing is known, the estimated TDOA provides information about the direction of the incoming wavefront relative to the microphone pair \cite{optimized_SRP_PHAT}. However, a single microphone pair generally does not provide sufficient information to uniquely determine the DOA. For a given TDOA, all possible source positions lie on a hyperbola in two-dimensional space or on a hyperboloid surface in three-dimensional space. Therefore, multiple spatially distributed microphones or multiple TDOA estimates obtained from different microphone configurations are required to resolve this ambiguity and estimate the DOA of the sound source \cite{hyperbolic_location}.


A common approach to estimating the TDOA from the microphone signals is based on cross-correlation. Cross-correlation quantifies the similarity between two time-dependent signals as a function of their relative delay. For sufficiently broadband signals and in the absence of strong noise or reflections, the location of the correlation peak can provide an accurate estimate of the TDOA. However, the estimation performance can degrade in the presence of noise, reverberation, or spectrally limited signals.
An extension of conventional cross-correlation is the generalized cross-correlation (GCC), in which the cross-power spectrum is weighted in the frequency domain before transformation to the time domain. A commonly used weighting scheme is the phase transform (PHAT), which normalizes the magnitude of the cross-power spectrum and thereby emphasizes phase information. Due to its relatively low computational complexity and robustness, particularly against spectral distortions, generalized cross-correlation with phase transform (GCC-PHAT) is widely used in numerous TDOA estimation approaches~\cite{GCC-PHAT, GCC_Performance}.

Another widely used method for DOA estimation is delay-and-sum beamforming. The method applies appropriate time delays to the signals of multiple microphones such that signals arriving from an assumed direction are temporally aligned and add constructively. In contrast, signals arriving from other directions only partially add coherently or interfere destructively, resulting in attenuation of their contributions to the beamformer output signal \cite{Beamforming}.
To determine the direction of a sound source using this approach, the beamformer output is calculated for a set of candidate angles. The angle at which the coherent summation reaches its maximum is considered to be the DOA of the sound source. The performance of the delay-and-sum beamformer strongly depends on the arrangement and number of microphones used. Uniform linear arrays are widely used because of their simple geometry and implementation, whereas circular or spherical arrays can provide directional sensitivity in both azimuth and elevation. Increasing the array aperture generally improves angular resolution, while increasing the number of microphones can improve the spatial selectivity and reduce side-lobe levels. However, excessively large microphone spacing can lead to spatial aliasing and ambiguities in the estimated direction \cite{Beamforming_for_DOA}.

Furthermore, direction estimation using delay-and-sum beamforming requires evaluating the beamformer output over a set of candidate directions, which can result in substantial computational effort, particularly when a fine angular search grid is used. In contrast, subspace-based methods can provide higher angular resolution and improved interference suppression under suitable conditions.

A common example of a subspace-based method is the Multiple Signal Classification (MUSIC) algorithm \cite{MUSIC}. Its fundamental concept is to decompose the spatial covariance matrix of the received microphone signals into two orthogonal subspaces: a signal subspace associated with the source signals and a noise subspace orthogonal to it. By applying an eigenvalue decomposition to the covariance matrix, the eigenvectors associated with the dominant eigenvalues form the signal subspace, whereas the eigenvectors associated with the remaining smaller eigenvalues form the noise subspace. MUSIC estimates the DOAs of sound sources by exploiting the orthogonality between the steering vectors, which describe the expected phase relationships between the microphone signals for a given DOA, and the noise subspace. MUSIC can provide high angular resolution and distinguish closely spaced sources under suitable conditions. However, the pseudospectrum is typically evaluated over a set of candidate directions, which can result in substantial computational effort when a fine angular search grid is used \cite{two_decades}.

Another subspace-based method for DOA estimation that avoids a spectral search is the Estimation of Signal Parameters via Rotational Invariance Techniques (ESPRIT) algorithm \cite{ESPRIT}. Like MUSIC, ESPRIT relies on a signal-subspace representation, typically obtained from the spatial covariance matrix. However, instead of evaluating a pseudospectrum over a set of candidate directions, ESPRIT exploits the rotational invariance between the signal subspaces of two identical subarrays related by a spatial translation. This relationship enables the source directions to be estimated directly from the signal subspace. By avoiding a spectral search, ESPRIT can reduce the computational effort compared with MUSIC. However, it requires an array geometry that provides the necessary translational invariance, which restricts the flexibility of sensor placement.

Since the proposed system aims to reduce the number of required microphones compared with conventional arrays, delay-and-sum beamforming and subspace-based methods are less suitable in their conventional forms. These approaches typically rely on predefined array geometries and multiple simultaneously available microphone signals. In contrast, TDOA-based methods can be applied to pairs of microphone signals and are therefore better suited to the proposed concept. Their application, however, requires precise knowledge of the microphone geometry associated with each TDOA estimate. Since the moving microphone does not follow a fixed or predefined trajectory, its position must be determined continuously during the measurement. The aim of this paper is therefore to develop and experimentally evaluate a two-channel microphone system with a time-varying geometry that exploits the motion of one microphone for DOA estimation, while fiber-optic position sensing is used to reconstruct the corresponding microphone trajectory.

\subsection{Fiber-Optic Shape Sensing}
\label{sec:FOPS}
The position of temporally and spatially varying microphones can be determined using fiber-optic shape sensing (FOSS), which enables continuous tracking of the shape of dynamic objects without external visual contact. In particular, this work is interested in the use of FOSS to determine the tip position of a flexible rod with a mounted microphone. In general, FOSS has been widely applied, for example, in structural health monitoring for aerospace applications~\cite{kissinger2022fibre} and in robotics~\cite{galloway2019fiber}, owing to advantages such as high differential strain resolution, fast response, large sensor multiplexing capability and flexibility~\cite{Review_2021fiber}. Here, FOSS enables the curvature, and consequently the shape, of a structure to be reconstructed when an optical fiber or fiber bundle is attached to it. Based on the measured curvature distribution, lateral displacements can then be determined by spatial integration.

In this work, a highly dynamic FOSS approach based on fiber segment interferometry (FSI) \cite{Kissinger_dynamicFOSS} is employed. In this method, interferometric sensing based on the range-resolved interferometry (RRI) method \cite{kissinger2015range} is used to integrate the fiber strain over extended fiber segments. Using three fibers arranged circumferentially around a rod, the bending of the rod in two dimensions relative to the rod base can be determined. While FSI-based FOSS allows in principle the use of several segments along each fiber to resolve complex rod shapes, a particular feature of the proposed approach is that for each of the three fiber, only a single fiber segment is employed. This is possible by assuming that the shape of the rod resembles the well-known cantilever mode shape. A considerable advantage of this approach is that no fiber Bragg grating reflector inscription is required in the three sensing fibers, enabling a very cost-effective sensor rod instrumented with standard telecom fiber pigtails. Furthermore, the measurements presented here have been conducted using the openly available interferometric system provided by the OpenRRI project at TU Ilmenau project~\cite{OpenRRI_2024}.

\section{PROPOSED SYSTEM}

\subsection{Theoretical Concept}

The proposed system is based on the evaluation of the TDOA between two microphones in combination with their geometric arrangement. As discussed in Section~\ref{sec:DOA}, a stationary setup consisting of only two microphones does not allow unambiguous DOA estimation in three-dimensional space. Instead, an estimated TDOA defines a hyperboloid representing all possible source positions consistent with that TDOA. The hyperboloid is described by the geometric relationship between the sound source and the microphone positions as 
\begin{equation}
\|\mathbf{s} - \mathbf{m}_1\| - \|\mathbf{s} - \mathbf{m}_2\| = c \cdot \Delta\tau_{12}
\label{eq:hyperboloid_tdoa}
\end{equation}
where $\mathbf{s}$ denotes the unknown position of the sound source, $\mathbf{m}_1$ and $\mathbf{m}_2$ denote the known microphone positions, $c = 343~\mathrm{m/s}$ is the speed of sound at $20\,^{\circ}\mathrm{C}$, and $\Delta\tau_{12}$ is the estimated TDOA between the microphone signals \cite{Conic_Approximation}. Consequently, a stationary two-microphone configuration does not provide sufficient information for unambiguous DOA estimation in terms of azimuth and elevation. This ambiguity can be resolved by combining multiple TDOA constraints obtained from different microphone configurations. In the proposed approach, these configurations are generated by moving one microphone relative to a stationary reference microphone, thereby creating a sequence of virtual microphone positions.

\begin{figure}[!h]
    \begin{center}
        \includegraphics[width=\columnwidth]{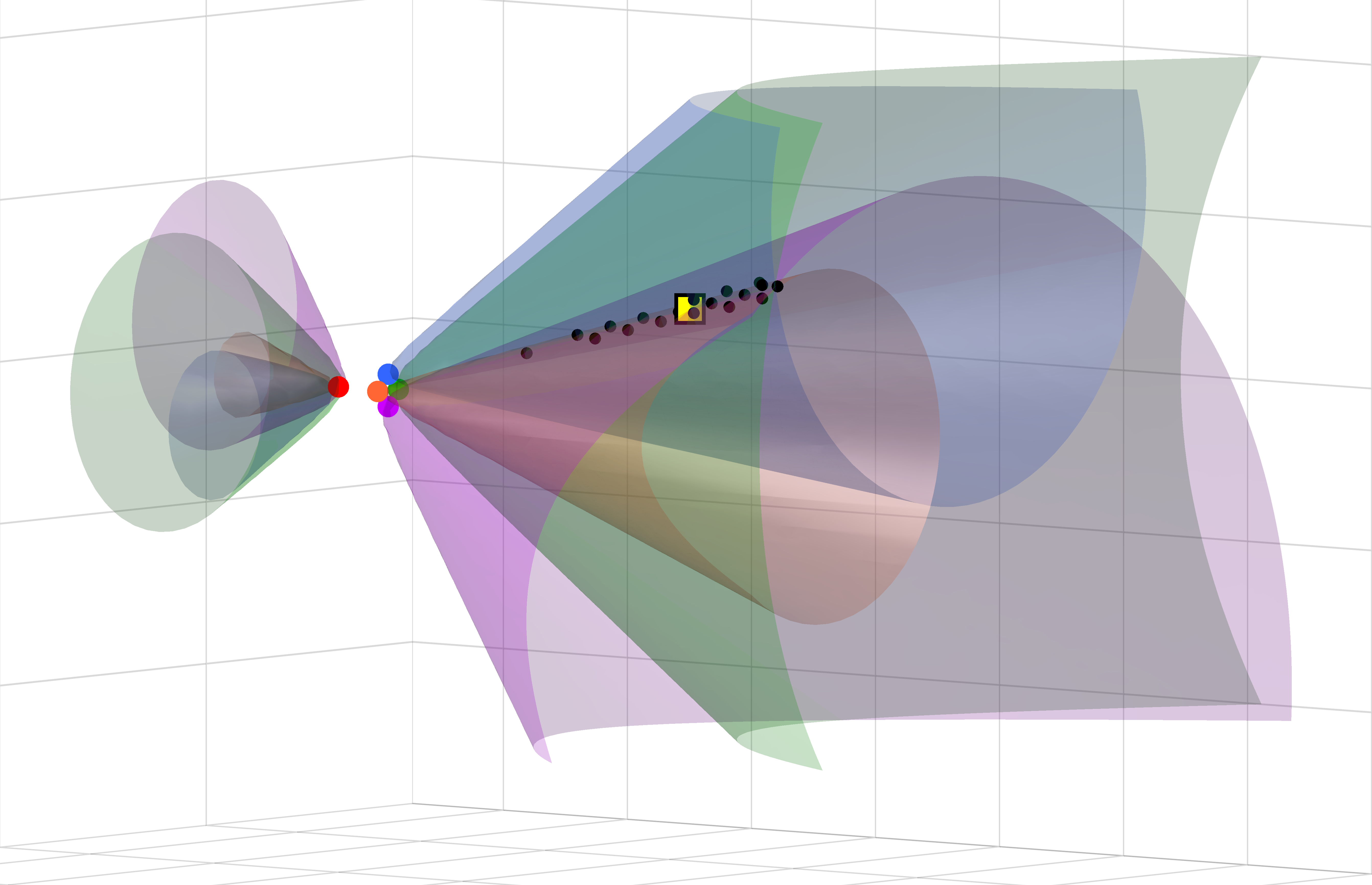}
    \end{center}
    \caption[\textbf{Illustration of the Hyperboloid Intersection Approach.}]{\textbf{Illustration of the hyperboloid intersection method.} 
    Each colored surface represents a hyperboloid defined by the TDOA between the stationary reference microphone and a virtual microphone position. Black dots indicate candidate grid points satisfying all hyperboloid constraints, and the yellow square marks the true source position. The coordinate-wise median of the retained points yields the estimated source position from which the DOA is derived.}
    \label{fig:Hyperboloid_Intersection}
\end{figure}

To realize this concept, the recorded signals are segmented into short time intervals during which the microphone position can be assumed to remain approximately constant, allowing each segment to be associated with a corresponding virtual microphone position. For each segment, the TDOA between the moving microphone and the stationary reference microphone is estimated. Together with the corresponding microphone configuration, each estimated TDOA defines a hyperboloid of possible sound source positions. By combining multiple hyperboloids and determining their common intersection region, the DOA of the sound source can be estimated in azimuth and elevation (Fig.~\ref{fig:Hyperboloid_Intersection}) \cite{Statistical_theory_of_passive_location_systems}.

Since the proposed approach uses non-predefined microphone motion to generate virtual microphone positions, the microphone trajectory must be determined throughout the measurement. As described in Section~\ref{sec:FOPS}, FOSS is used to reconstruct the time-varying shape of the supporting structure. The optical fibers are attached to the supporting structure, and the moving microphone is mounted at a known position on it. When the structure is excited, the microphone follows its motion. Based on the reconstructed shape of the supporting structure and the known microphone mounting position, the microphone position can be determined over time. The resulting trajectory is subsequently used to derive the virtual microphone positions required for TDOA-based DOA estimation.

\subsection{Simulation}

To evaluate the feasibility of the proposed concept and investigate the influence of relevant system parameters, the approach was analyzed in a simulation environment. The simulation combines a simplified mechanical model of the supporting structure with an acoustic free-field propagation model. Coupling both submodels enables the generation of synthetic microphone signals with known source and microphone positions for the development and evaluation of the signal processing chain.
The first submodel describes the motion of the oscillating supporting structure, which is approximated by a damped harmonic oscillator. This simplified representation is used to generate a time-varying microphone trajectory rather than to reproduce the exact structural dynamics of the experimental setup. Assuming small angular displacements, the motion is described by
\begin{equation}
    \ddot{\theta}(t) + 2\gamma \dot{\theta}(t) + \omega_0^2 \theta(t) = 0\,,
\end{equation}
where $\theta(t)$ denotes the angular displacement, $\gamma$ is the damping coefficient, and $\omega_0 = 2\pi f_0$ is the undamped natural angular frequency \cite{MIT_Damped_Oscillations}. The parameter $L$ defines the effective length of the supporting structure and determines the spatial extent of the resulting microphone trajectory. The oscillator model is applied independently to the two lateral spatial axes, resulting in the angular displacements $\theta_y(t)$ and $\theta_z(t)$. The model parameters, including $L$, the oscillation frequency, damping coefficient, and initial conditions, can be varied to investigate different microphone motion characteristics.

The second submodel describes the acoustic propagation from the sound source to the microphones. The simulation assumes free-field conditions and considers only the direct sound component. The sound source is modeled as a point source, resulting in a propagation delay and distance-dependent attenuation at each microphone position. For a microphone position $\mathbf{r}_{\mathrm{mic}}(t)$ and a source position $\mathbf{r}_{\mathrm{src}}$, the propagation distance $d(t)$ and the corresponding propagation delay $\tau(t)$ are given by
\begin{equation}
    d(t) = \|\mathbf{r}_{\mathrm{src}} - \mathbf{r}_{\mathrm{mic}}(t)\|
    \qquad \text{and} \qquad
    \tau(t) = \frac{d(t)}{c}\,.
\end{equation}
The simulated microphone signal is then described by
\begin{equation}
    s_{\mathrm{mic}}(t) = \frac{1}{d(t)}\,s_{\mathrm{source}}\!\left(t-\tau(t)\right)\,,
\end{equation}
where $s_{\mathrm{source}}(t)$ denotes the digital source signal and $c$ the speed of sound. In the discrete implementation with sampling frequency $f_s$, the time-varying propagation delay is realized as a fractional-sample delay. Since the delay generally does not correspond to an integer number of samples, the delayed signal is obtained by linear interpolation between adjacent samples. The source signal is selected according to the requirements of the respective simulation scenario and specified during initialization.

\subsection{Data Processing Pipeline}

Based on the simulated data, a signal processing pipeline was developed consisting of three main steps: data preprocessing, TDOA estimation between the microphone signals, and the determination of the hyperboloid intersection region used for DOA estimation.

First, the simulated microphone signals and the corresponding position data of the moving microphone are divided into consecutive windows of equal length. Each window constitutes a separate processing unit from which one DOA estimate is obtained. The window length determines the temporal resolution of the DOA estimation and therefore affects the ability to track time-varying source conditions. Based on preliminary investigations, a window length of $T_{\mathrm{window}}=0.5\,\mathrm{s}$ was selected. Each processing window is subsequently divided into overlapping blocks used for TDOA estimation.
The block length $N_{\mathrm{block}}$ affects the balance between the amount of signal information available for TDOA estimation and the temporal variation of the microphone geometry within a block. Short blocks may contain insufficient signal information for robust TDOA estimation, whereas longer blocks increase the microphone displacement occurring during the analyzed signal segment. For each block, the TDOA between the moving and stationary microphone signals is estimated using GCC-PHAT. The corresponding microphone trajectory samples are evaluated over the same time interval. Since the microphone position varies within each block, the arithmetic mean of these positions is used as the representative virtual microphone position $\mathbf{m}_{\mathrm{virt}}$.

In the next step, the blocks used for constructing the hyperboloids are selected. To obtain an unambiguous estimate of the sound source direction from the intersection of multiple hyperboloid constraints, several TDOA estimates associated with sufficiently distinct virtual microphone configurations are required in three-dimensional space \cite{jin2018robust}. Therefore, the number of selected blocks $N_{\mathrm{select}}$ is systematically varied in the subsequent parameter study to investigate its influence on the accuracy and robustness of the DOA estimation and to identify suitable parameter settings. In this process, the blocks are selected so that they are evenly distributed across the processing window. After block selection, the resulting set of $N_{\mathrm{select}}$ measurements consists of the estimated TDOAs and the corresponding virtual microphone positions.

For estimating the hyperboloid intersection region, a three-dimensional Cartesian search-grid, $\mathbf{x}=(x,y,z)^\mathrm{T}$, is defined.

The grid boundaries and sampling density were selected empirically based on preliminary trials to provide a suitable balance between spatial resolution and computational effort. 
For a virtual microphone position $\mathbf{m}_\text{virt}$ and the stationary reference microphone position $\mathbf{m}_\text{stat}$, the theoretical path difference is given by
\begin{equation}
\Delta d(\mathbf{x}) = \|\mathbf{x} - \mathbf{m}_\text{virt}\| - \|\mathbf{x} - \mathbf{m}_\text{stat}\|
\end{equation}
and compared with the estimated path difference
\begin{equation}
\Delta d_\text{est} = c \cdot \Delta\tau_\text{est}\
\end{equation}
which is obtained from the TDOA estimation. A grid point is considered consistent if the deviation between the theoretical and estimated path differences remains within a tolerance $\Delta_\mathrm{tol}$:
\begin{equation}
    \big| \Delta d(\mathbf{x}) - \Delta d_\text{est} \big| < \Delta_\mathrm{tol}\,.
\end{equation}
This tolerance is required because the discrete search grid generally does not contain points that exactly satisfy the hyperboloid constraints. Its value is therefore related to the spatial resolution of the search grid and was selected empirically together with the search-grid parameters.
This condition is evaluated for all selected blocks, and only grid points satisfying all TDOA constraints are retained. To reduce the influence of spatially distributed candidate points, the coordinate-wise median of the retained grid-point coordinates is calculated. The resulting position estimate is converted into azimuth and elevation relative to the coordinate system of the microphone system. This procedure is repeated for each processing window, providing a sequence of DOA estimates over the analyzed signal duration.

\subsection{Parameter Study}
For the practical implementation of the proposed concept, the influence of the main system parameters was investigated. These include parameters of the motion model, such as the supporting structure length, oscillation frequency, damping coefficient, and initial displacement, signal-processing parameters including block length and hop size, and parameters of the hyperboloid intersection algorithm. The parameters were systematically varied in the simulation environment and evaluated with respect to DOA estimation accuracy and robustness.

The parameter study showed that a supporting structure length of $L=0.3\,\mathrm{m}$ in combination with an initial displacement of $\theta_0=0.2\,\mathrm{rad}$ and an undamped natural frequency of $f_0=4\,\mathrm{Hz}$ provided favorable conditions for the subsequent DOA estimation. For signal processing, a block length of $N_{\mathrm{block}}=2048$ samples and a hop size of $N_{\mathrm{hop}}=1024$ samples provided a suitable balance between estimation performance and temporal resolution. For the hyperboloid-based DOA estimation, the highest accuracy and robustness within the investigated parameter range were obtained with $N_{\mathrm{select}}=6$ blocks.

The simulation results provide an initial basis for selecting suitable system parameters for the practical implementation of the proposed concept. However, the identified parameter values should be regarded as preliminary because the simplified motion model does not reproduce the actual structural dynamics of the experimental system and the free-field simulation does not represent the full complexity of real acoustic environments.

\subsection{Technical Implementation}

\begin{figure}[!h]
    \begin{center}
        \includegraphics[width=\columnwidth]{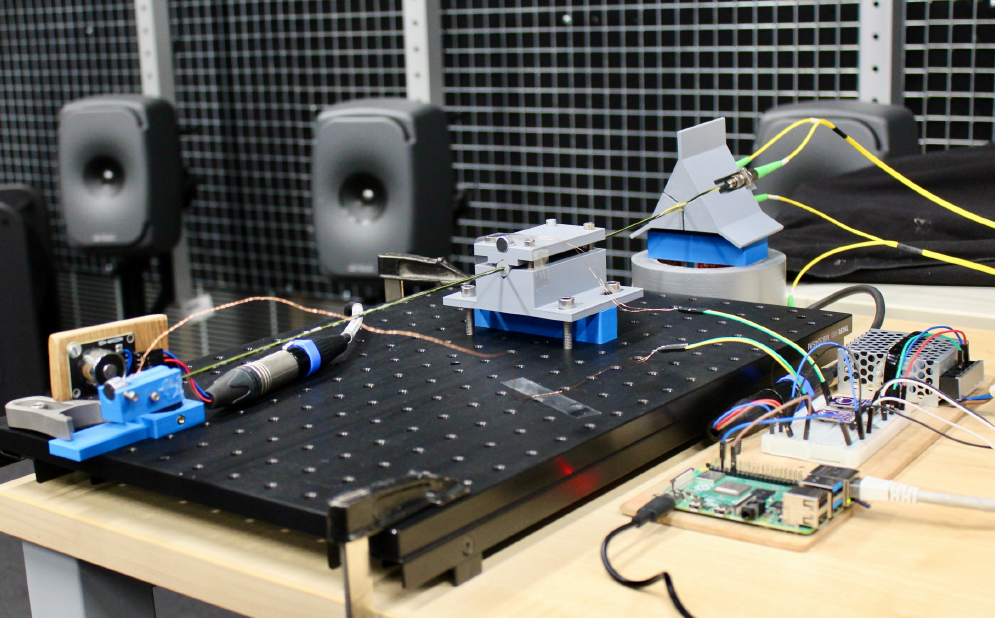}
    \end{center}
   \caption[\textbf{Image of the Implemented Technical System}]{\textbf{Image of the implemented system}. 
   The supporting structure and the excitation mechanism are mounted on a baseplate using mounting fixtures. The microcontroller, microphone preamplifier, and power supply are installed on a separate plate.
   }
    \label{fig:System}
\end{figure}

For the practical implementation of the proposed concept, an experimental system integrating the components required for audio acquisition and microphone trajectory reconstruction was developed (Fig.~\ref{fig:System}). 
While the proposed DOA algorithm constitutes the core of the signal processing, its experimental realization requires several interconnected subsystems. These include the mechanical supporting structure and its clamping mechanism, the fiber-optic interferometric sensing system for reconstructing the microphone trajectory, the acoustic acquisition system for recording the microphone signals, and the excitation system used to initiate the oscillatory motion of the supporting structure and the measurement sequence.

The mechanical structure forms the physical basis of the proposed system and generates the oscillatory motion that defines the geometry of the virtual microphone array. A carbon fiber rod with a diameter of $d_{rod}=2\,\mathrm{mm}$ was selected as the supporting structure. Owing to its cylindrical geometry, the rod does not exhibit a geometrically preferred bending direction and can therefore support oscillatory motion in different spatial planes. This enables the generation of a two-dimensional microphone trajectory, which is required to obtain sufficiently distinct virtual microphone configurations for the subsequent DOA estimation. Based on the parameter study, a free rod length of $L=0.3\,\mathrm{m}$ was selected.

The oscillatory motion of the supporting structure is initiated using its elastic restoring force in combination with an electromagnet. A ferromagnetic disk is attached to the microphone mount at the tip of the rod and serves as the magnetic counterpart. Before each measurement, the rod is manually deflected to a predefined initial deflection and held in place magnetically. When the magnetic field is switched off, the rod is released and undergoes a freely decaying oscillation. This mechanism provides a reproducible initialization of the microphone trajectory.

The acoustic acquisition system was designed to provide synchronized microphone signals while minimizing the mechanical influence on the supporting structure. Compact, low-mass electret condenser microphone capsules with external preamplifiers were therefore selected. The microphone signals were digitized simultaneously using an audio interface at a sampling frequency of $f_{\mathrm{s}}=192\,\mathrm{kHz}$.

\begin{figure}[h!]
\centering
\includegraphics[width=82mm]{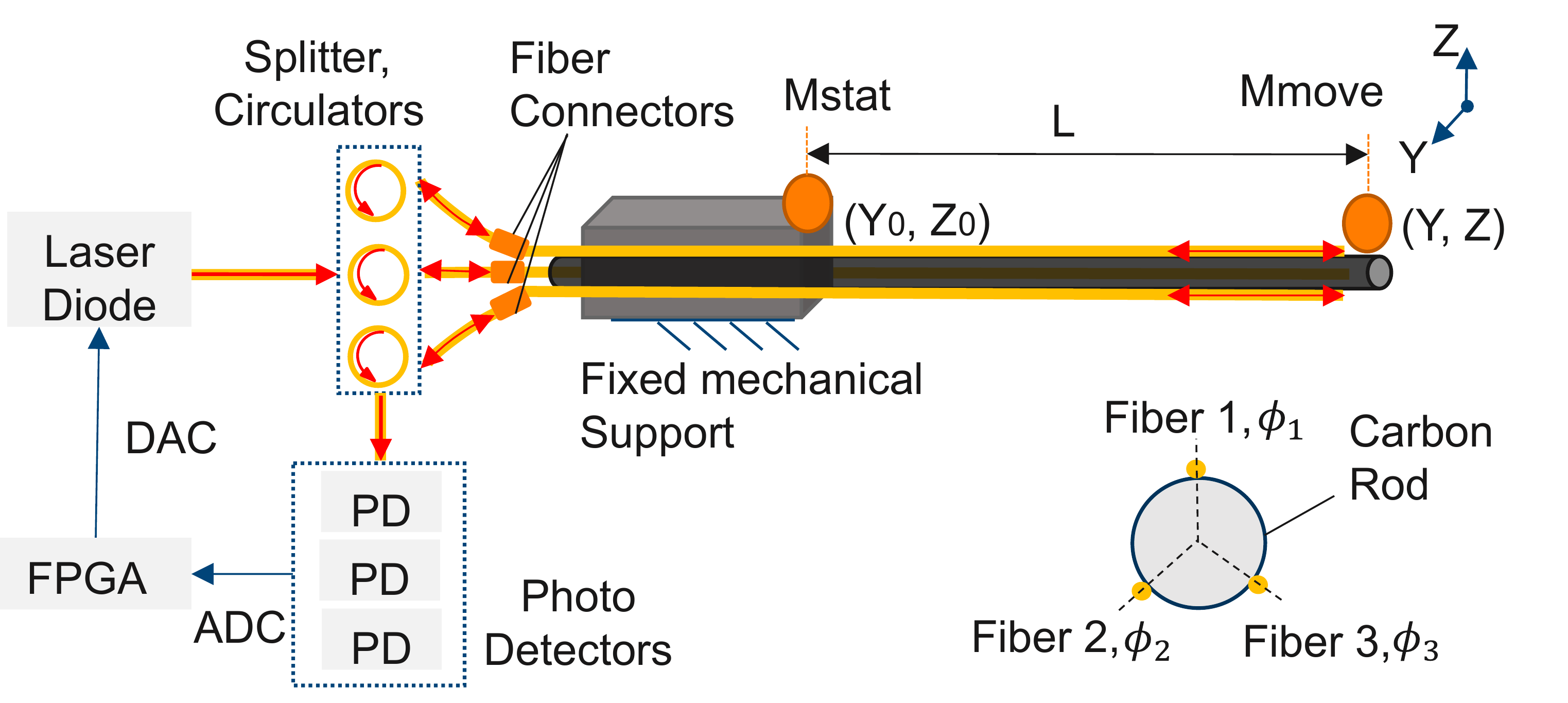}
\caption{\textbf{Principle of the fiber segment interferometry configuration used for lateral displacement sensing.}}
\label{fig:Setup}
\end{figure}

Fig.~\ref{fig:Setup} shows the optical setup of the FOSS system used in the technical implementation, which consists of a laser diode emitting light at $\lambda_0 = 1550\,\mathrm{nm}$. For the RRI processing, the laser injection current is modulated at 40~kHz by an FPGA-based Red Pitaya data acquisition board using the openly available OpenRRI implementation~\cite{OpenRRI_2024}. The laser light is then split into three single-mode optical fibers, each with a length of 60~cm, which are bonded to the surface of a carbon rod clamped on one side.

The fibers are arranged with equal angular spacing ($120^\circ$) around the rod. In each fiber, the desired interference occurs on the one side by the Fresnel reflection of the air gap within the mismatched FC/PC and FC/APC combination and on the other side by the Fresnel reflection at the cleaved fiber tip. This enables a very cost-effective realisation of the sensor fibers using standard telecom fiber pigtails. The interference signals are then guided by fiber optic circulators to three photodetectors, and the resulting phase signals from the interference are demodulated using RRI and subsequently converted into tip displacement measurements \cite{Kissinger_dynamicFOSS}.

As discussed, the present work assumes a simple cantilever model and in this model the tip displacement varies linearly with the slope angle change between the start and end of the free cantilever. Therefore, at least for small displacements, the measured interferometic phase signals are directly proportional to the desired tip displacements \cite{Kissinger_dynamicFOSS}. However, in the present setup, precise alignment of the fiber cores and accurate determination of other fiber sensor parameters, such as the fiber-core group refractive index and fiber strain sensitivity factor, are challenging. Therefore, in this work, the tip displacement $d$ needs to be calibrated and can be generically expressed as:
\begin{equation}
\label{eq:displacement_phase_relation}
d = K \Delta \phi .
\end{equation}
\noindent Here, $K$ is the calibration coefficient of the fiber-optic displacement sensor and $\Delta\phi$ is the differential interferometric phase measured. The sensor phase response is calibrated by applying known purely horizontal and vertical displacements, $\Delta y$ and $\Delta z$, respectively, to the cantilever tip using a linear stage on the optical table:
\begin{equation}
K_y = \frac{\Delta y}{\Delta\phi_y} \quad \text{and} \quad K_z = \frac{\Delta z}{\Delta\phi_z}\,.
\end{equation}
whereby the directional projections of the phase values are calculated from the differential phases:
\begin{equation}
\phi_y(t) = \phi_3(t) - \phi_2(t) \quad \text{and}
\end{equation}
\begin{equation}
\phi_z(t) = \frac{1}{2}\big(\phi_3(t) - \phi_1(t) + \phi_2(t) - \phi_1(t)\big)\,.
\end{equation}
For the subsequent measurement, the conversion is applied in reverse. The lateral positions are obtained by multiplying the measured phase changes by the previously determined calibration coefficients:
\begin{equation}
y(t) = K_y \cdot \phi_y(t) \quad \text{and} \quad z(t) = K_z \cdot \phi_z(t)\,,
\end{equation}

This provides the time-resolved displacement of the rod tip relative to the starting shape of the measurement, where the starting shape is well defined by the magnetic holding mechanism. Subsequently, the lateral displacement is calculated and transformed into absolute coordinates within the coordinate system of the measurement setup. In this way, the DOA processing pipeline receives the microphone trajectory relative to the stationary microphone, allowing the virtual microphone positions to be determined.

The position and audio data require temporal alignment for joint evaluation. Both acquisitions are controlled by a Python script, but since the systems do not share a common hardware clock, the recorded data streams were manually aligned during post-processing.
For the experimental data, the processing pipeline developed in the simulation-based parameter study is retained with modifications required for the practical measurements. During block selection, additional quality and stability criteria are applied to retain only plausible blocks for the subsequent DOA estimation. The resulting processing chain forms the basis for the experimental evaluation of the proposed approach.

\begin{figure*}[!b]
\centering
\includegraphics[width=1\textwidth]{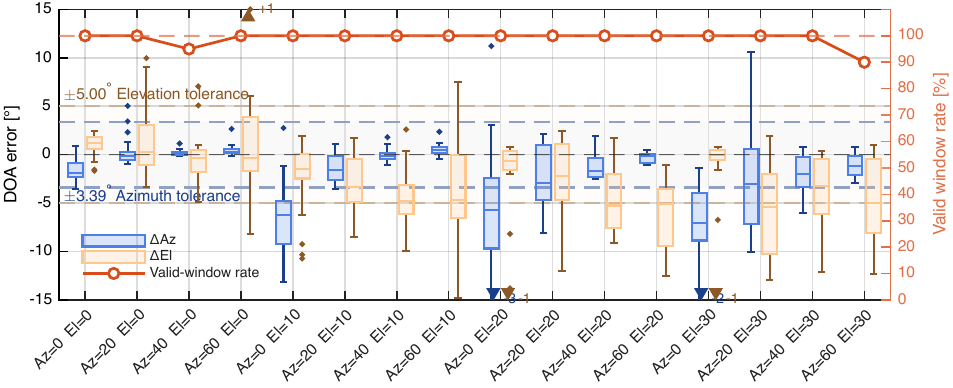}
   \caption{\textbf{Error distribution and valid-window rate for different sound source positions}. The valid-window rate remains close to 100\,\%. The median azimuth error is within tolerance for 13 of the 16 source positions, while the median elevation error is within tolerance for 12 positions. Across the valid processing windows, 71\,\% of the elevation estimates are within tolerance. Errors increase for frontal sources at higher elevations and for elevation estimates at larger azimuth angles.}
    \label{fig:Mean_total_error}
\end{figure*}

\section{EVALUATION}

The objective of the experimental evaluation is to demonstrate the feasibility of the proposed system and to assess its performance in terms of DOA estimation accuracy and robustness. The evaluation further investigates how selected system and measurement conditions affect the estimation performance. For this purpose, four factors are investigated: sound source position, source signal characteristics, sound source distance and microphone motion. In addition, the performance of the proposed system is evaluated for a moving sound source to investigate its behavior under dynamic source conditions.
All measurements were conducted in a low-reflection laboratory environment under approximately constant temperature and humidity conditions. A loudspeaker with a coaxial driver configuration was employed as the acoustic source. The loudspeaker position was determined using an optical tracking system, which also provided the reference trajectory for the evaluation involving a moving sound source.

To ensure consistent and comparable measurements across all experimental trials, an identical measurement protocol was applied throughout the evaluation. The protocol included the alignment and calibration of the initial position of the moving microphone, adjustment of the loudspeaker output level at the stationary reference microphone, and execution of the measurement. Each measurement had a total duration of $t_{\mathrm{meas}}=11\,\mathrm{s}$, of which $t_{\mathrm{data}}=10\,\mathrm{s}$ were used for data processing. The remaining $t_{\mathrm{buff}}=1\,\mathrm{s}$ was excluded from the analysis to account for synchronization offsets and transient effects at the beginning and end of the measurement. With a processing window length of $T_{\mathrm{window}}=0.5\,\mathrm{s}$, the recorded data yield 20 DOA estimates per measurement. The selected measurement duration also allows the influence of the decaying oscillation of the supporting structure on the DOA estimation performance to be evaluated.

Two performance metrics are used throughout the evaluation. The angular deviation between the estimated and reference DOA in azimuth and elevation is used as the accuracy metric, whereas robustness is quantified by the valid-window rate, defined as the fraction of processing windows for which the algorithm calculates an intersection region and provides a DOA estimate. To account for the finite spatial extent of the loudspeaker, an angular tolerance is introduced for evaluating the estimation error. The tolerance is derived from the physical dimensions of the loudspeaker and its distance from the system reference point and is recalculated for each investigated source distance. Estimated DOAs with angular deviations within the corresponding tolerance are classified as correct.

\subsection{Influence of the Sound Source Position}

To investigate the influence of the sound source position on system performance, 16 measurements were performed with the sound source placed at different positions relative to the microphone system. The source positions were selected to yield azimuth angles of $\varphi_{\mathrm{source}} \in \{0, 20, 40, 60\}\,^\circ$ and elevation angles of $\theta_{\mathrm{source}} \in \{0, 10, 20, 30\}\,^\circ$ relative to the coordinate system of the microphone setup. Negative azimuth and elevation angles were excluded because the electromagnet and clamping mechanism would obstruct the direct propagation path between the loudspeaker and the microphones, potentially introducing shadowing and diffraction effects. The sound source distance was kept constant at $d_{\mathrm{source}}=2\,\mathrm{m}$, while the loudspeaker orientation relative to the room remained unchanged. White noise was used as the source signal at a sound pressure level of $70\,\mathrm{dB(A)}$ at the stationary reference microphone.

Fig.~\ref{fig:Mean_total_error} shows the azimuth and elevation error distributions together with the valid-window rates for all 16 source positions. For the azimuth estimation, the median error lies within the tolerance of $\varphi_{\mathrm{tol}}=\pm3.39^\circ$ for 13 of the 16 source positions, while approximately $76\,\%$ of the valid processing windows yield estimates within this tolerance. The highest azimuth accuracy is observed at larger source azimuth angles, with more than $98\,\%$ of the valid processing windows remaining within tolerance for $\varphi_{\mathrm{source}}\geq40^\circ$. In contrast, azimuth estimation deteriorates near $\varphi_{\mathrm{source}}=0^\circ$ as the elevation angle increases, with predominantly negative errors and increased dispersion.

The elevation estimation exhibits a similar overall accuracy, with median errors within the tolerance of $\theta_{\mathrm{tol}}=\pm5^\circ$ for 12 of the 16 source positions and approximately $71\,\%$ of the valid processing windows within this range. The highest elevation accuracy is observed for $\varphi_{\mathrm{source}}=0^\circ$, where more than $91\,\%$ of the valid processing windows remain within the tolerance. As the azimuth angle increases, the elevation errors become more dispersed and the fraction of estimates within tolerance decreases. Overall, the results indicate a direction-dependent estimation performance. Azimuth estimation is most accurate at larger azimuth angles but deteriorates near the frontal direction as elevation increases. In contrast, elevation estimation performs best near the frontal direction and becomes less accurate and more variable at larger azimuth angles.

\subsection{Influence of the Sound Source Distance}

The influence of the sound source distance was investigated using six measurements at fixed source angles of $\varphi_{\mathrm{source}}=20^\circ$ and $\theta_{\mathrm{source}}=10^\circ$, with sound source distances of $d_{\mathrm{source}}\in\{1,2,3,4,5,6\}\,\mathrm{m}$. A valid-window rate of $100\,\%$ was obtained at all investigated distances. This consistently high rate may be facilitated by the low-reflection measurement environment, which reduces reflections that could adversely affect TDOA estimation. The mean azimuth error ranges from $-1.3^\circ$ to $-0.3^\circ$ and remains within the corresponding tolerance over the entire distance range. Elevation estimation shows slightly larger mean errors, ranging from $-4^\circ$ to $-1^\circ$, which is consistent with the lower elevation accuracy observed in the preceding source-position analysis. The error dispersion also increases with sound source distance, with standard deviations of approximately $3^\circ$ and $4^\circ$ for $d_{\mathrm{source}}\geq3\,\mathrm{m}$. The predominantly negative errors in both azimuth and elevation may indicate a possible systematic bias in the measurement and estimation chain. However, the hypothesis that the DOA estimation error decreases with increasing sound source distance due to the reduced angular extent of the loudspeaker is not supported by the measurements.

\subsection{Influence of the Microphone Motion }

\begin{figure}[!h]
    \begin{center}
        \includegraphics[width=1\columnwidth]{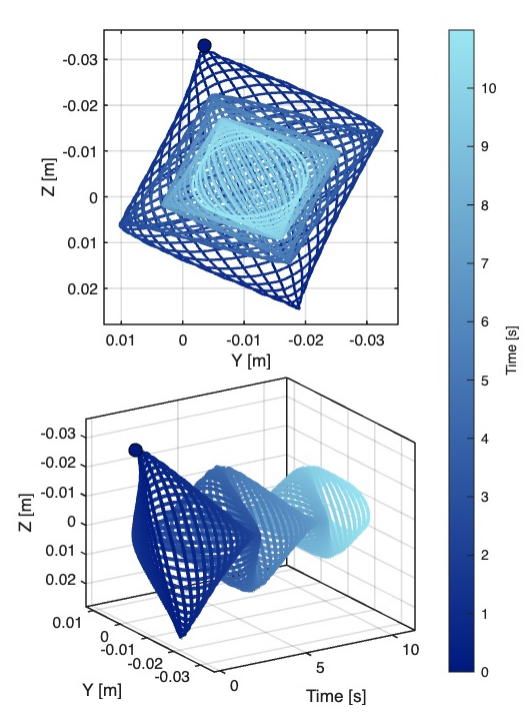}
    \end{center}
    \caption[\textbf{Microphone trajectory.}]{\textbf{Microphone trajectory measured by the FOSS system}. The shown trajectory over a 10 s period, starting after release marked by the blue dot, follows a narrow elliptical pattern whose extent decreases over time due to oscillation damping, while its orientation varies periodically in the $y$-$z$ plane.}
    \label{fig:Movement}
\end{figure}

The microphone motion is investigated as an internal system parameter that directly determines the geometry of the virtual microphone array. The measured trajectories are highly consistent across all measurements, indicating reproducible motion of the supporting structure. 
The mean maximum displacements are $d_x=2\,\mathrm{mm}$, $d_y=54\,\mathrm{mm}$, and $d_z=42.6\,\mathrm{mm}$.
The parameter study identified an initial displacement angle of $\theta_0=0.2\,\mathrm{rad}$, corresponding to a maximum displacement of $d_{\mathrm{sim}}\approx56.9\,\mathrm{mm}$ for the implemented supporting structure length of $L=0.30\,\mathrm{m}$. The measured displacement in the y-direction is of a similar magnitude to the simulated maximum displacement. In contrast, the measured oscillation frequency was approximately $f_{\mathrm{meas}}\approx12.1\,\mathrm{Hz}$, whereas the simulation used an undamped natural frequency of $f_{\mathrm{0}}\approx4\,\mathrm{Hz}$. The microphone trajectory, shown in Fig.~\ref{fig:Movement}, follows a narrow elliptical pattern whose orientation varies periodically in the $y$-$z$ plane.

\begin{figure*}[b]
\centering
\includegraphics[width=1\textwidth]{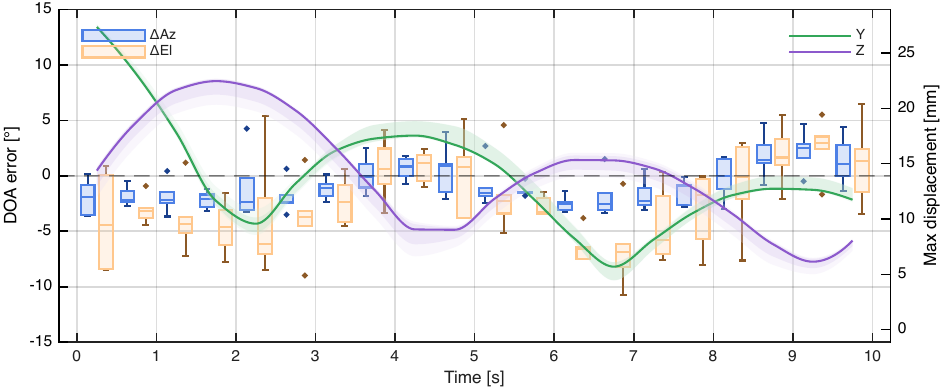}
   \caption[\textbf{DOA error distribution in relation to microphone displacement.}]{\textbf{DOA error distribution in relation to microphone displacement.} The figure shows the distribution of DOA errors for measurements $\varphi_{\mathrm{source}}=20^\circ$, $\theta_{\mathrm{source}}=10^\circ$ and varying distances with $d_{\mathrm{source}}\in\{1,2,3,4,5,6\}\,\mathrm{m}$ and the corresponding maximum microphone deflections in the x-, y- and z-directions (median, IQR, 5th--95th percentiles). Larger displacements in the z-direction are associated with predominantly negative DOA errors, whereas smaller displacements tend to be associated with positive errors.}
   \label{fig:Auswertung_Bewegung}
\end{figure*}

Periodic variations in the DOA estimation error were observed over the duration of the measurements. Since the sound source remains stationary within each measurement, these variations may be related to the periodically changing microphone motion. To investigate this relationship, six measurements with fixed source angles of $\varphi_{\mathrm{source}}=20^\circ$ and $\theta_{\mathrm{source}}=10^\circ$ and sound source distances of $d_{\mathrm{source}}\in\{1,2,3,4,5,6\}\,\mathrm{m}$ were analyzed. For each processing window, the azimuth and elevation errors were compared with the maximum microphone displacement along the three spatial axes (Fig.~\ref{fig:Auswertung_Bewegung}). 

The results indicate a relationship between the DOA estimation error and the axis-specific microphone displacement. In particular, decreasing displacement along the y-axis is associated with increasing errors in both azimuth and elevation, with a more pronounced trend for elevation. For the z-axis, larger displacements are associated with predominantly negative errors, whereas smaller displacements tend to be associated with positive errors. 

Overall, the results suggest that the microphone trajectory affects the DOA estimation performance. However, the displacements along the three spatial axes are coupled and cannot be assessed independently. Moreover, the observed relationships are specific to the investigated microphone motion and source direction. A more detailed assessment would therefore require additional measurements in which the microphone motion is varied systematically under otherwise comparable conditions.

\subsection{Influence of the Sound Source Signal}

The influence of source signal characteristics on the DOA estimation performance was investigated using seven signals at a fixed sound source position and distance. The signals comprised white noise at sound pressure levels of $L_{\mathrm{source}}\in\{70,60,50\}\,\mathrm{dB(A)}$, high-pass-filtered white noise with cutoff frequencies of $f_{\mathrm{cut}}\in\{500,3000,6000\}\,\mathrm{Hz}$, and a speech signal. The high-pass cutoff frequencies were selected based on the crossover frequencies of the loudspeaker.

Variations in sound pressure level and spectral content resulted in only minor differences in the DOA estimation performance. The consistently high valid-window rates across the investigated sound pressure levels may be facilitated by the low-noise measurement environment and the resulting high signal-to-noise ratio. At $f_{\mathrm{cut}}=6000\,\mathrm{Hz}$, a slight reduction in the valid-window rate was observed, suggesting lower robustness when the available signal bandwidth is reduced. The hypothesis that adapting the source spectrum to the loudspeaker crossover frequencies improves DOA estimation accuracy was therefore not supported by the measurements.

For the speech signal, mean azimuth and elevation errors of $-1.8^\circ$ and $-3.8^\circ$, respectively, remain within the corresponding tolerance ranges, while the valid-window rate decreases to $70\,\%$. This reduction may be related to the time-varying spectral content of speech, which can reduce the number of signal segments suitable for robust TDOA estimation.

\subsection{Performance with a Moving Sound Source}

Finally, the performance of the proposed system was evaluated for a moving sound source to investigate its behavior under dynamic source conditions. Source motion continuously changes the geometric relationship between the source and the microphones and therefore the TDOA associated with each microphone configuration. This variation can affect the consistency of the resulting hyperboloid constraints and their common intersection region.
For the experiment, the loudspeaker was moved laterally along the $y$-axis with a maximum velocity of $v_{\mathrm{source}}\approx0.64\,\mathrm{m/s}$. Its trajectory was recorded using the optical tracking system and converted into the corresponding reference DOA for evaluation of the estimates. White noise was used as the source signal.

With a processing window length of $T_{\mathrm{window}}=0.5\,\mathrm{s}$, a valid-window rate of $95\,\%$ was obtained, indicating that the algorithm returned valid DOA estimates for most processing windows despite the time-varying source geometry.
The azimuth error exhibits a median of approximately $-1^\circ$ and an interquartile range of $2.8^\circ$. In contrast, the elevation error has a median of approximately $-3.8^\circ$ and an interquartile range of $10.5^\circ$. The results therefore show higher accuracy and lower error variability in azimuth than in elevation under the investigated dynamic conditions.
Overall, the results demonstrate the feasibility of DOA estimation for a moving sound source under the investigated conditions. However, further measurements with different source trajectories, velocities, and signal characteristics are required to assess the performance under a broader range of dynamic conditions.

\section{CONCLUSION}

This paper presented and experimentally demonstrated the feasibility of a temporally and spatially variable microphone array in which fiber-optic interferometric sensing is used to determine virtual microphone positions. By reconstructing the trajectory of a moving microphone relative to a stationary reference microphone, the proposed approach provides a time-varying array geometry for TDOA-based DOA estimation without requiring a predefined microphone trajectory.

Across the experiments, azimuth estimation generally showed higher accuracy and lower error variability than elevation estimation. Predominantly negative errors under several measurement conditions further suggest a possible systematic bias. Its origin cannot be identified from the current experiments and may be related to the experimental setup, trajectory reconstruction, or assumptions and parameters within the estimation method. 

Moreover, the evaluation was performed in a controlled low-reflection environment with a single active sound source, which limits conclusions regarding more complex acoustic scenarios. Future work should therefore investigate systematically varied microphone motion, improved trajectory reconstruction, and more robust TDOA and hyperboloid-intersection methods, followed by evaluations in reflective environments, with multiple simultaneously active sources, and under more diverse dynamic source conditions.

The experiments also highlight the trade-off introduced by the trajectory-sensing approach. Fiber-optic sensing enables microphone motion to be reconstructed without requiring a predefined trajectory, but the current implementation involves substantial calibration and post-processing effort. 
Further developments could extend the approach to multiple moving microphones, where the position and orientation of multiple fiber segments are measured using FSI in FBG-inscribed fibers \cite{Kissinger_dynamicFOSS}, enabling richer time-varying array geometries and additional spatial constraints.

Compared with the method used in this study, multiple microphones could provide greater spatial redundancy and improved robustness, particularly for signals with a low signal-to-noise ratio or in reflective environments, while potentially reducing the required observation time. A larger number of simultaneously recording microphones would also enable alternative DOA estimation methods, including subspace-based approaches such as MUSIC. In contrast to a static array, the additional microphone motion could provide temporally varying and potentially noncoplanar array geometries, thereby increasing spatial diversity and reducing directional ambiguities.

The concept may additionally be extended to application-specific sensor arrangements, including microphones suspended below drones or integrated into flexible, textile, foldable, or rollable structures. Such configurations could exploit structural motion directly and potentially provide larger effective apertures or improved separation from application-specific noise sources.

\section*{ACKNOWLEDGMENT}
The authors (JW, TK) gratefully acknowledge the support by the Deutsche Forschungsgemeinschaft, Germany (DFG) in the scope of the Priority Programmes SPP 2433 “Metrology on flying platforms” (Grant Number 541029557) at Technische Universität Ilmenau, Germany. The authors further thank the Ilmenau Interactive Immersive Technologies Centre (I3TC) for providing access to the VR laboratory in which the measurements were conducted.

\balance

\bibliographystyle{unsrt} 
\bibliography{Bibliography}

\end{document}